\documentclass[%
 reprint,
 amsmath,amssymb,
 aps,prd
]{revtex4-1}

\usepackage{graphicx}
\usepackage{dcolumn}
\usepackage{bm}
\usepackage{hyperref}
\usepackage{tabularx}
\usepackage{amssymb}
\usepackage{amsthm}
\usepackage{amsmath}
\usepackage{caption}
\usepackage{setspace}
\usepackage{enumitem}
\usepackage{textcomp}
\usepackage[section]{placeins}
\usepackage[utf8]{inputenc}
\usepackage{wrapfig}
\usepackage{epstopdf}
\usepackage[toc,page]{appendix}
\usepackage{enumitem}
\usepackage{xcolor}

\newcommand{\beq}{\begin{equation}}
\newcommand{\eeq}{\end{equation}}
\newcommand{\Mtop}{m_{\textrm{top}}}

\begin{document}
\preprint{APS/123-QED}

\title{Asymptotic safety, the Higgs mass and beyond the Standard Model physics}

\author{Jan H. Kwapisz${}^{1,2}$}%
 \affiliation{${}^1$Institute of Theoretical Physics,
Department of Physics, University of Warsaw, ul. Pasteura 5, 02-093 Warsaw, Poland}
 \affiliation{${}^2$Max-Planck-Institut f\"ur Gravitationsphysik (Albert-Einstein-Institut), Am Mühlenberg 1 D-14476 Potsdam, Germany}

\email{Jan.Kwapisz@fuw.edu.pl}

\date{\today}


\begin{abstract}
There are many hints that gravity is asymptotically safe. The inclusion of gravitational corrections can result in the ultraviolet fundamental Standard Model and constrain the Higgs mass to take the smallest value such that electroweak vacuum is stable. Taking into account the current top quark mass measurements this value is $m_H \approx 130$ GeV.\\
This article considers the predictions of the Higgs mass in two minimal Beyond Standard Model scenarios, where the stability is improved. One is the sterile quark axion model, while the other is the $U(1)_{B-L}$ gauge symmetry model introducing a new massive $Z'$ gauge boson. The inclusion of $Z'$ boson gives the  correct prediction for this mass, while inclusion of sterile quark(s) gives only a slight effect.\\ 
Also a new, gravitational solution to the strong CP problem is discussed.

\end{abstract}

    


\maketitle
\section{Introduction}
The couplings of the physical models change with scale, and there are two sources of this scaling. The first, classical scaling is due to canonical dimensionality of the operators. The theory, which is classically scale invariant possesses dimensionless couplings only in the action. This is indeed the case for the Standard Model with zero bare Higgs mass, so called Conformal Standard Model \cite{MEISSNER2007}. The other source of scaling is caused by the quantum effects, which can spoil classical scale invariance and provide the generation of scale due to radiative corrections. In particular the Coleman-Weinberg mechanism generates masses in this pattern \cite{PhysRevD.7.1888}. In quantum field theories the change (``running") of couplings with energy scale is described by renormalisation group equations 
\beq
\mu \frac{\partial}{\partial \mu} g_i(\mu)=\beta_i(\{g_j\}).
\eeq
Such a general equation can have various possible behaviours for $\mu \to +\infty$, yet only some of them makes the theory predictable up to the infinite energies. In the simplest case the couplings reach the fixed point ($\forall_i \beta_i(\{g_j)\})=0$) and the running stops, making the theory scale invariant on the quantum level. However, this is not only possibility, since the coupling can also be attracted to a higher dimensional structure \cite{Gukov:2016tnp}, like a limit cycle (see for example \cite{Dawid:2017ahd,Moroz:2009nm} for a limit cycle behaviour in $1/r^2$ potential) or a chaotic attractor. Such theories can also be UV fundamental, yet they are not scale invariant. On the other hand, scale invariance seems to play some fundamental role in the construction of the quantum gravity theory(ies), see \cite{THooft:2015jcw,tHooft:2017avq, Rachwal:2018gwu,Wetterich:2019qzx}, so in this article we restrict to the fixed point case. The fixed point can be at zero (Gaussian fixed point), making the theory asymptotically free. Alternatively it can reach some non-zero value (non-Gaussian fixed point / residual interaction). We call such theory asymptotically safe. Steven Weinberg hypothesised that gravity possesses an interacting fixed point \cite{Weinberglectures,Weinberg}. This issue was studied in \cite{Gastmans:1977ad,Christensen:1978sc,Smolin:1981rm}, where the calculations were done by means of $\epsilon$ expansion in the vicinity of $2$ dimensions. However, in general such fixed points cannot be considered by means of ordinary perturbation theory, where one does expansion of the theory around the fixed point at zero. The study of such fixed points requires other, non-perturbative treatment. \\
The functional renormalisation group (FRG) is one of the tools which can be used. In the FRG approach one studies the evolution of the effective average action $\Gamma_k$, which is an quantum effective action, where all the interactions with momenta lower than $k$ are integrated out. The $\Gamma_k$ interpolates between the classical action $S_{\Lambda}$ at the UV scale $\Lambda$ and the full quantum effective action $\Gamma = \Gamma_{k=0}$. The evolution of $\Gamma_k$ is given by the Wetterich equation \cite{Wetterich:1992yh,Morris:1993qb,Morris:1994ie}. Using this approach the gravitational fixed points were found (for Euclidean signature), see \cite{Reuter:1996cp,Dou:1997fg}. Moreover, the gravitational corrections to the matter beta functions can be calculated and they alter the UV running of the matter couplings. Despite the fact that the asymptotic safety programme for quantum gravity is far from being finished, see \cite{Eichhorn:2018yfc,percaccibook}, yet it seems to be a very promising way to quantise gravity, not only because of its simplicity, but also due to its rich particle physics phenomenology, which can be tested. In particular two years before the discovery of the Higgs boson, its mass was calculated in \cite{ShaposWetterich} as $126 \pm \textrm{few}$ GeV. However, the authors took the top quark mass smaller than the current observed value. In this article we repeat this calculation and investigate the possible sources of disparity between current experimental measurements and the theoretical predictions. \\
However the asymptotic safety programme has more predictions in case of particle physics. For example, since the top Yukawa coupling is close to the upper bound in the basin of attraction, hence if it runs to the interacting fixed point, then it is also predictable \cite{Eichhorntop}. In such scenario the difference between the top and the bottom quark masses \cite{Eichhorn:2019yzm} can also be predicted. Moreover the fine structure constant in Grand Unified Theories can also be predicted \cite{Eichhorn:2017muy}. These results are promising, however the results for the Higgs mass calculated for the top interacting fixed point scenario \cite{Eichhorntop} gives $m_H \approx 132$ GeV. The authors stress that the \emph{results arise in a truncation of the RG flow that is limited to the surmised leading-order effects of quantum gravity on matter} \cite{Eichhorntop}. This might be the case, see \cite{Gies2017,Eichhorn:2015kea,Loebbert:2015eea}, however the Planck suppressed couplings doesn't affect the running in the IR \cite{Branchina:2013jra,Branchina:2014usa}. The $\lambda(\mu)$ becomes negative at $10^{10}$ GeV and in this article we explore another possibility, namely that addition of Beyond Standard Model fields results in the correct predictions for the Higgs mass. The fact that most of the problems of the Standard Model can be solved at $\sim 1$ TeV scale \cite{Shapo2,Lewandowski2017,SBCS,SBCS2} supports that view and the new physics should affect the prediction of the Higgs mass. In particular we analyse two scenarios: addition the $Z'$ boson, which is related to the B-anomalies and addition of sterile quarks.  

 
 \section{Calculation of the Higgs Boson mass in the Standard Model}
 In this paragraph we revaluate the calculations done in \cite{ShaposWetterich} concerning the calculation of the Higgs mass. The Higgs part of Standard Model Lagrangian is given by:
\beq
\mathcal{L}_{\textrm{Higgs}}=(D_{\mu}H^{\dagger}D^{\mu}H) - \lambda \left((H^{\dagger}H)^2-v^2\right)^2,
\eeq 
 where $v_H\approx246.22$ GeV. On the tree level one has:
\beq
m_H^2=2\lambda  v^2,
\eeq
and the radiative corrections are $\mathcal{O}(1)$ GeV. The one-loop beta functions (where $\hat{\beta}_{SM} = 16\pi^2\beta_{SM}$) in the $\overline{MS}$-scheme are:
\beq
\begin{array}{ll}
\hat{\beta}_{g_1}& =  \frac{41}{6} g_1^3, \textrm{   }  \hat{\beta}_{g_2} = - \frac{19}{6} g_2^3, \textrm{   }  \hat{\beta}_{g_3} = - 7 g_3^3,\\
\hat{\beta}_{y_t} &=  y_t\left(\frac{9}{2} y_t^2 - 8 g_3^2 - \frac{9}{4} g_2^2 - \frac{17}{12} g_1^2\right), \\
\hat{\beta}_{\lambda_1} &=  24 \lambda_1^2 - 3 \lambda_1\left( 3 g_2^2 + g_1^2 - 4y_t^2\right)\\ 
&+\frac{9}{8} g_2^4 + \frac{3}{4} g_2^2 g_1^2 + \frac{3}{8} g_1^4 - 6 y_t^4,
\end{array}
\eeq
 where $g_1$, $g_2$, $g_3$ are $U(1)$, $SU(2)$, $SU(3)$ Standard Model gauge couplings respectively and $y_t$ is the top Yukawa coupling. The two-loop beta functions, we have used in our calculations, are given in \cite{Machacek:1984zw,Arason:1991ic}. The gravitational corrections \cite{Robinson:2005fj,ShaposWetterich,Eichhorn,Pawlowski:2018ixd} to the beta functions are in the leading order:
\beq
\label{Gravitationalcorrections}
\beta^{\textrm{grav}}_i (g_i,\mu)= \frac{a_i}{8\pi}\frac{\mu^2}{M_P^2+2\xi_0\mu^2}g_i,
\eeq
where $M_P=2.4 \times 10^{18}$ GeV is the low energy Planck mass, $\xi_0$ is related to the gravitational fixed point and depends on the matter content, see Eq. (\ref{GN}). For the Standard Model one has $\xi_0 \approx 0.024$ and $a_{\lambda}=+3$, $a_{y_t}=-0.5$, $a_{g_i}= -1$. Depending on the sign of $a_i$ one gets repelling / attracting fixed point at zero for a given coupling. If one demands that all of the matter couplings to be asymptotically free then the ones with the repelling fixed points becomes predictable and the one with attracting fixed points have to be inside the basin of attraction, otherwise they will diverge \cite{ShaposWetterich}. Since $a_{\lambda}=+3$, then Higgs self coupling has a repelling fixed point at zero, and becomes a prediction of a theory rather than being a free parameter. On the two-loop level and for $y_t=g_1=g_2=g_3=0$ one has:
\beq
 \beta_{\lambda}(\mu) =\frac{1}{16\pi^2} \left( 24\lambda^2- \frac{312}{16\pi^2}  \lambda^3\right) + \frac{a_{\lambda}}{8\pi} \frac{\mu^2}{M_P^2+2\xi_0\mu^2} \lambda,
\eeq
which has the following fixed points: $\lambda = 0$ (repeller), $\lambda \approx 21$ (attractor), $\lambda \approx -9.36$ (attractor). 
 For this reason two basins of attraction are separated by the single trajectory going to the repelling fixed point. The numerical calculations confirm that if at any scale below Planck scale $\lambda(\mu) <0$ then it drops to the non-perturbative fixed point. If one assumes that $\lambda$ has to stay in the perturbative region, then necessarily one gets $\lambda(\mu) \geq 0$ at all scales. Furthermore in order to avoid the attractor in the positive domain one should assume that (again confirmed by numerics at two loop level):
\beq
\label{lambda1}
\lambda = \min\{ \overline{\lambda}: \forall_{\mu} \overline{\lambda}(\mu)\geq 0, \overline{\lambda}(M_P) \approx 0 \textrm{ and } \beta_{\bar{\lambda}},(M_P) \approx 0\}
\eeq
which agrees with the arguments of the authors of \cite{ShaposWetterich}. Then one needs stable EW vacuum in order to predict the Higgs mass in the line of \cite{ShaposWetterich}. This reasoning explains also the Multiple Point Principle postulated in \cite{Bennett:1993pj}. According to this principle there are two vacuum states with about the same energy density, one at electroweak scale and one at the Planck scale, which can be used to predict the SM couplings \cite{Bennett:1993pj,Froggatt:1995rt,Froggatt:2003tu}. It has also interesting cosmological consequences \cite{Sidharth:2018dds}. Let us note that the requirement Eq.~(\ref{lambda1}) is actually stronger than the EW stability, since the stability can be affected by Planck suppressed operators, while positivity of $\lambda$ is not affected by Planck physics \cite{Branchina:2013jra,Branchina:2014usa}. \\
The current calculations of running of $\lambda(\mu)$ shows that $\lambda(\mu)$ drops to negative values at roughly $10^{10}$ GeV \cite{Buttazzo:2013uya,Bezrukov:2012sa}, making the vacuum metastable. Also the situation is similar if one takes into account the non-minimal $H^{\dagger}H R$ term \cite{Bezrukov:2014ipa}. However, it was suggested in \cite{Bednyakov:2015sca} that our vacuum can be stable for some space of parameters. Yet for central values of $\Mtop$ and $m_H$ the vacuum is metastable with the estimation of the lower stability bound is $m_H > (129.6 \pm 1.5) $ GeV \cite{Buttazzo:2013uya,Bezrukov:2012sa}. On the other hand from the experimental point of view the Higgs mass is constrained as: $m_H=125.18 \pm 0.18$ GeV \cite{PhysRevD.98.030001}, which corresponds to $\lambda(\Mtop) = 0.127823\pm0.000367$ in $\overline{MS}-$scheme for one-loop matching conditions \cite{Sirlin,Machacek:1984zw,Arason:1991ic} (and $\lambda=0.12924\pm 0.00037$ at the tree level), where we have taken into account the uncertainties in the measurements of the top quark $\Mtop = 173.0 \pm 0.4$ \cite{PhysRevD.98.030001}. Hence the stability of EW vacuum, assumed in \cite{ShaposWetterich} is in contradistinction with the measured Higgs value. Yet this stability bound is close enough to the experimental value of the Higgs mass, that one can hope that a slight extensions of the SM can bring it to the correct value. \\
To obtain the predictions for $\lambda$ we do the two-loop running of the $g_1,g_2,g_3, y_{t}, \lambda$ with gravitational corrections and search for optimal $\lambda$ for given set of $g_1,g_2,g_3,y_t$, such that $\lambda\geq 0$ and there are no Landau Poles ($\lambda$ does not end in the non-perturbative region). Then given $\lambda$ one can recover the Higgs mass via matching relations (let us note that we treat $v$ as given from experiment).\\
In our analysis we take one-loop-matched parameters as \cite{Buttazzo:2013uya}: $g_1 (\Mtop) = 0.35940$, $g_2(\Mtop) = 0.64754$, $g_3(\Mtop)=1.18823$, and we scan over one-loop matched $y_t$ for various experimentally viable $\Mtop$, giving $y_{top}(\Mtop) = 0.94759\pm 0.0022$, which is slightly lower than the central value obtained in \cite{Buttazzo:2013uya}. As a result we get $\lambda= 0.15102 \pm 0.00158$ giving $m_H \approx 135$ GeV at one-loop and $\lambda =0.13866 \pm 0.00218$  at two-loops (the uncertainties are due to the $y_t$ coupling) and $m_H \approx 130.5$ GeV. Actually the two-loop result is close to the stability bound of the Higgs mass, which means high degree of accuracy. \\
We have checked that if one takes the bottom quark and the taon into account, it changes the predictions for $m_H$ less then $1$ KeV, which is far below the theoretical and experimental accuracy. This can be expected since $y_b(\Mtop) \approx 0.015$ \cite{Bednyakov:2016onn}. Due to metastability of vacuum we see that it is necessary to introduce the beyond Standard Model operators in order for Higgs mass to be predicted in the asymptotic safety paradigm at the correct experimental value.
\section{Beyond Standard Model}
\subsection{Gravitation constraints}
First let us constrain the possible additional matter content. We have:
\beq
\begin{array}{lcr}
\label{GN}
\xi_0 = \frac{1}{16 \pi G_N^{\ast}}& \textrm{ and }  &G_{N}^{\ast} \approx -\frac{12 \pi}{N_S+2N_D - 3N_V -46},
\end{array}
\eeq
where $N_S, N_D, N_V$ are the number of scalars, fermions and vector particles respectively. We know that $G_N(\Mtop) \geq 0$ and the running of $G_N$ cannot change the sign of $G_N$ \cite{Reuter:1996cp}. Then we have $G_N^{\ast}\geq0$. So the Beyond Standard Model Theories which extend the Standard Model broadly can be incompatible with the asymptotic safety paradigm. For example such theories are MSSM \cite{PhysRevD.24.1681} and some of the GUTs \cite{BURAS197866,IBANEZ1981439}. It seems \cite{percaccibook,Wetterich:2019zdo} that asymptotic safety prefers the minimal extensions of the Standard Model. One should note that these effects can possibly be truncations artefacts and can disappear after taking into account more operators. On the other hand the actual change of $\xi_0$ due to small modifications of the SM doesn't alter the predictions at observable level. \\
\subsection{Models}
As we have said the prediction of $\lambda$ depends highly on the initial value of top Yukawa coupling. It also strongly depends on the running of $y_t$. So changing this running alters the prediction of $\lambda$ from asymptotic safety. In this paragraph we shall discuss two extensions of the Standard Model, where $\beta_{y_t}$ is slightly changed, because to predict correct value of $\lambda$ it seems that only a minor effect is required. In both models we extend the Higgs sector by an additional complex scalar singlet under $SU(3)_c, SU(2)_L\times U(1)_Y$ gauge groups \cite{Higgsportal}:
\beq
\mathcal{L}_{\textrm{scalar}} = (D_{\mu}H)^{\dagger}(D^{\mu}H) + (\partial_{\mu}\phi^{\ast}\partial^{\mu}\phi) - V(H,\phi), 
\eeq 
\begin{eqnarray}
\label{Vphi}
V(H,\phi) = -m^2_1 H^{\dagger}H - m_2^2\phi^{\star}\phi + \lambda_1(H^{\dagger}H)^2 \nonumber \\
+ \lambda_2(\phi^{\star}\phi)^2  + 2\lambda_3(H^{\dagger}H)\phi^{\star}\phi.
\end{eqnarray}
Often one also includes right handed neutrinos coupled to $\phi$ \cite{Lewandowski2017,Shaposhnikov:2006xi,Drewes:2013gca}, yet they won't be relevant to our discussion. The inclusion of portal interaction stabilises the vacuum \cite{MEISSNER2007,Lewandowski2017} (also with inclusion of higher order operators \cite{Eichhorn:2014qka}), yet in our further analysis we shall put $\lambda_3=0$. So there will be no portal stabilisation effect. This is in line with the asymptotic safety analysis of such models \cite{Eichhorn:2017als}, where one needs $\lambda_3=0$ at all scales. \\
\subsubsection{Model I}
In the Model I the global, not anomalous SM group  $U(1)_{B-L}$, related to the baryon minus lepton $(B - L)$ number, is gauged and
a new Gauge boson $B_{\mu}'$ \cite{Langacker:2008yv,Chankowski:2006jk,Basso:2011hn} is introduced. Then the covariant derivatives receives an additional contribution: $D_{\mu} \to D_{\mu} + i(\tilde{g}Y + g_1' Y_{B-L})B_{\mu}'$, where $Y$  is the hyper-charge and $Y_{B-L}$ is the (B-L)-charge. The $B_{\mu}'$ boson becomes massive due to the non-zero vacuum expectation value of $\phi$, and the $\tilde{g}$ describes the mixing between $Z$ and $Z'$ after spontaneous symmetry breaking. Following \cite{Basso:2011hn} we analyse the ``pure'' $B-L$ model by assuming that there is no tree level mixing between Z bosons ($\tilde{g}(\Mtop) =0$), which is supported by the current data \cite{PhysRevD.98.030001}, which it might be spoiled by radiative corrections. This model is also supported experimentally, since it is the most popular way of explaining so called $B$-anomalies \cite{Aaij:2015oid,Aaij:2017vbb,Sierra:2015fma,Altmannshofer:2014rta}, which are the observed inconsistencies of the SM with experimental data in the bottom quark decays. The new terms in the beta functions at the one-loop level are (for $\tilde{g}=0$):
\beq
\begin{array}{lcr}
\hat{\beta}_{g_1'}= 12 g_1'^3, &\hat{\beta}_{y_t}= \hat{\beta}_{y_t}^{\textrm{SM}}  -\frac{2}{3}y_t g_1'^2\theta(\mu-M_{Z'}).
\end{array}
\eeq
The mass of the $Z'$ boson is restricted to be: $\frac{M_{Z'}}{g_1'}> 7 $ TeV or $M_{Z'}\leq 2\Mtop$ \cite{Chiang:2014yva}.  
\subsubsection{Model II}
The Model II is inspired by KSVZ axion \cite{Kim,Vainshtein} and includes new sterile (EW singlet) quarks $Q_i$ charged under $U(1)_{PQ}$ coupled to new scalar:
\beq
\begin{array}{lcl}
\mathcal{L} &=& \mathcal{L}_{\textrm{fermions}}+\mathcal{L}_{\textrm{Y}}+  \mathcal{L}_{\textrm{gauge}}
+ \mathcal{L}_{\textrm{scalar}}\\ & + & \sum_{i=1}^n\left(\bar{Q}_i D_{\mu}\gamma^{\mu} Q_i - y_Q\phi\bar{Q}_iQ_i +\textbf{h.c.}\right),
\end{array}
\eeq
where we assume that Yukawa matrix $\mathbf{y}_Q$ to be diagonal and the quarks acquire masses $M_i = y_Q v_{\phi}/\sqrt{2}$. The ``phase'' of $\phi$ is called the axion particle and becomes massive due to instanton effects. This model was proposed to solve the Strong CP problem \cite{Ellis:1978hq} by spontaneous symmetry breaking of $U(1)_{PQ}$ \cite{PhysRevLett.38.1440}. As a side comment let us note that asymptotic safety gives a possible explanation to the strong CP problem without axions. The strong CP-violation consists of two terms $\theta_{QCD} = \theta_{topological} + \textrm{arg det} M_u M_d$ and in principle $\textrm{arg det} M_u M_d$ should give much bigger contribution to the strong CP-violation. By considering the gravitational corrections, the following reasoning can, at least partially, explain the smallness of the strong CP-violation effect. Namely, in the case of $\textrm{arg det} M_u M_d$ there is no running till at least 7-loops \cite{Ellis:1978hq}. Despite the fact that the gravitational corrections Eq.~(\ref{Gravitationalcorrections}) are extremely small, yet they can overtake the dynamics even in the IR and drop $\textrm{arg det} M_u M_d$ to zero, since the matter contributions are $\mathcal{O} \left((\textrm{arg det} M_u M_d)^{17}\right)$. In order for gravitational contributions to be dominant one needs $ 0.01 \gtrsim  \textrm{arg det} M_u M_d$, which is far beyond the experimental bounds. However, this argument requires a more detailed analysis. \\
 Even with $\lambda_3=0$ the running of $g_3$ is affected by the inclusion of the heavy quarks:
\beq
16\pi^2\frac{d g_3}{d \log \mu} =\hat{\beta}(g_3) \to \hat{\beta}(g_3)+\frac{2}{3}\sum_{i=1}^n \theta(\mu-M_{Q_i}) g_3^3,
\eeq
which in turn alters the running of $y_t$. Let us note that if there are many such quarks, then even the asymptotic freedom of the QCD can be spoiled. However, in our analysis we focus on addition of one or two sterile quarks into the SM. 

\subsection{Calculations}
\begin{figure}[h!]
\centering
\includegraphics[width=1\linewidth]{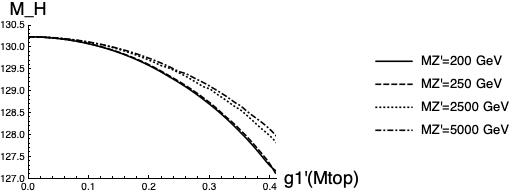}
\caption{Higgs mass for various $g_1'$ and $M_{Z'}$, $y_t=0.94759$}
\label{fig1}
\end{figure}

\begin{figure}[h!]
\centering
\includegraphics[width=1\linewidth]{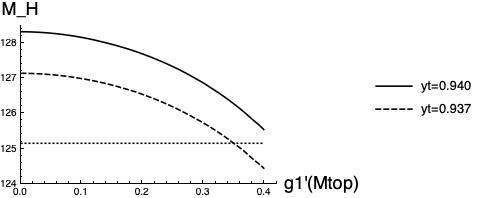}
\caption{Higgs mass for $M_{Z'}=200$ GeV, higher loop matching}
\label{fig3}
\end{figure}

\begin{figure}[h!]
\centering
\includegraphics[width=1\linewidth]{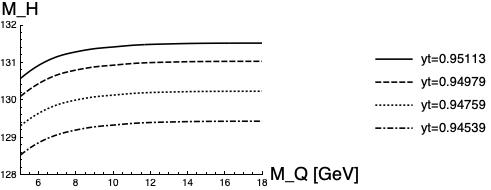}
\caption{Optimised $\lambda(\Mtop)$ for various $M_Q$(in logarithmic scale) and $y_t$}
\label{fig2}
\end{figure}

At first let us note that both models agree with the condition given by Eq.~(\ref{GN}), with $\xi_1 =0.02$ and $\xi_2 =0.023$ accordingly. Asymptotic safety requirement gives restrictions on the couplings of new degrees of freedom. In case of \textbf{Model I}, one gets that $g_1'(\Mtop) \in [0.0,0.4]$ (with $a_{g_1'}, a_{\tilde{g}}=-1$). On the other hand the minimal mass for the sterile quark from \textbf{Model II} is $m_{Q} \sim 100$ TeV, otherwise the running becomes unstable. Furthermore if one includes one more quark, then its mass is of the order of $10^6$ TeV giving a huge hierarchy, which seems to be very unnatural. Since then we shall restrict ourselves to one heavy, sterile quark. Below, on Fig.~[\ref{fig1}] we present the calculations for \textbf{Model I} using the two loop beta functions for the couplings, see \cite{Das:2015nwk,delAguila:1988jz,Luo:2002iq}.
\FloatBarrier
\FloatBarrier
As we can see on the plot for large $g_1'$ and small $M_{Z'}$ the Higgs mass is getting close to the experimental value. For $y_t=0.9539$ one even gets $m_H \approx 126$ GeV. If we use more precise formulas for two-loop matching \cite{Buttazzo:2013uya} then we get that the central value $y_t\approx0.94$. For this central value and for $g_1'>0.3$, $m_{Z'} <2\Mtop$ Higgs mass is $m_H = 125 \pm 1.5$ GeV depending on the exact parameters. On Fig.~\ref{fig3} we show this dependence using two-loop matched $y_t=0.94$ and two-loop matched $y_t=0.936$ with higher order QCD corrections.
\FloatBarrier
\FloatBarrier
 The \textbf{Model I} satisfies criteria of EW vacuum stability and is of experimental and theoretical interest. The asymptotic safety predicts hat the mass of the new gauge boson should be small, which can be verified experimentally. Furthermore our argument is confirmed by the fact that for certain space of parameters the EW vacuum in the B-L extension is stable \cite{Das:2015nwk,DiChiara:2014wha,Coriano:2015sea,Wang:2015sxe}. The effect of introducing $Z'$ boson can be even more significant if the Higgs boson is also charged under $U_{B-L}$, see \cite{DiChiara:2014wha}. Yet in such models $Z'$ is highly constrained observationally with $M_{Z'}> 3$ TeV. Moreover if one relax the condition $\lambda_3(\mu) =0$, which is the case in more general $B-L$ models, then one immediately gets the stability of EW vacuum \cite{PhysRevD.99.115029,Latosinski2015,Lewandowski2017} and hence correct Higgs mass.\\
Furthermore we have checked that the $\tilde{g}$ corrections and inclusion of right handed neutrinos, botom quark and taon gives the contributions which are negligible. For instance taking into account the $y_N$ right handed neutrino Yukawa coupling two loop contribution \cite{Das:2015nwk} gives the difference of $10$ MeV between $y_N=0.0$ and $y_N=0.44$ situation. 
\\
In the case of \textbf{Model II} we perform the full two-loop analysis, with one quark $Q$ and with masses in range $m_Q\in(10^5 - 10^{18})$ GeV. The results are shown below on Fig.~[\ref{fig2}].
\FloatBarrier
\FloatBarrier
For \textbf{Model II} the new degrees of freedom influence the running of $\lambda$ much less than in \textbf{Model I}. The change in predicted Higgs mass at the one-loop matching is of order $1.5$ GeV downwards. There are two reasons for that. First of all addition of $Q$ changes only running of $g_3$, which in turn changes the running of $y_t$ and has only slight effect on $\lambda$. Secondly the new degrees of freedom are constrained to have mass far beyond the EW scale, while the $Z'$ mass isn't constrained that much both theoretically and observationally. We can conclude that inclusion of additional sterile quarks cannot drop the Higgs mass to the correct value. Yet maybe \textbf{Model II} combined with \textbf{Model I} can give the correct Higgs mass. \\
\section{Discussion}
There are a few issues which require separate discussion. First of all, to obtain the running of the considered couplings one can solve the full Wetterich equation, see for example \cite{Eichhorntop,Eichhorn:2018whv,Eichhorn:2017egq,Wetterich:2017aoy}. While Wetterich equation  is exact, however it is very difficult (or even impossible) to be solved, because one has to take into account all of the operators which coincide with the symmetries. Moreover one has to choose the cutoff, which is arbitrary \cite{Litim:2001up,percaccibook}. So in order to reproduce the correct perturbative results one has to take into account many higher order operators and choose the proper cutoff. As a state of the art the current FRG calculations match the usual results at the one loop level and the leading contributions to running at two-loop level \cite{Wetterich:2017aoy} for pure gauge theories. Moreover the gravitational corrections are ambiguous due to gauge dependence, for example the prediction of top Mass ranges from $130$ to $171$ GeV only due to this effect \cite{Eichhorntop}. For the sake of phenomenology we decided to use the loop expansion and the EFT gravitational corrections \cite{Robinson:2005fj} supplemented with the gravitational fixed point calculated with the FRG techniques \cite{Wetterich:2019zdo}. Furthermore it seems that these two approaches give similar results (compare the fixed point of top Yukawa coupling in \cite{ShaposWetterich} and \cite{Eichhorntop}).\\
One can also argue that $a_{i}$ \cite{ShaposWetterich,Zanusso:2009bs,Robinson:2005fj} are not calculated to high accuracy, making the whole calculation very sensitive to those parameters, hence not-reliable. This is indeed the case for non-Gaussian fixed point making the prediction of upper bound for top quark mass sensitive to new physics \cite{Eichhorntop}. Yet in the case of Gaussian fixed point the existence of attractive / repelling fixed point at zero is much more vital than actual value of $a_{\lambda}$ due to the stability argument. \\
In our analysis we use the $\overline{MS}$ beta functions and parameters. For this scheme we cannot use the Appelquist-Carazzone theorem \cite{PhysRevD.11.2856} and we rely on the effective field theory approach \cite{Weinberg:1980wa}. In this approach one has to take into account the threshold effects \cite{Ross:1978wt,Hall:1980kf}. In case of \textbf{Model II} the these threshold effects at two-loop precision change the prediction of Higgs mass by $5$ MeV, which is far below both experimental and theoretical accuracy. On the other hand for \textbf{Model I}, due to relative small mass of $Z'$, one should fit the observables to the new set of $\overline{MS}$ couplings and then do the running, see discussion in \cite{Chankowski:2006jk}. Otherwise one can use mixed on-shell / $\overline{MS}$ scheme \cite{Chankowski:2006jk}. Both procedures are beyond the scope of this article and are left for future work. As a matter of fact this theoretical uncertainties are superseded by much bigger experimental ones in the top and Higgs mass measurements (the fact that $g_1'$ is relevant parameter also contributes to this uncertainty).\\
Finally the $\xi_0$ depends not only on the matter content, but also on the gravity sector. For example in unimodular gravity \cite{Eichhorn:2013xr} it has slightly different value, yet the effect on Higgs mass is negligible (in naive calculations one gets $\mathcal{O}(1$ MeV)), yet it might be interesting to test it in the future. On the other hand there are other more fundamental modifications of gravity, like massive gravity \cite{deRham:2014zqa} or Horava gravity \cite{Horava:2009uw}, and their fixed point structure might be completely different. Then with the right theoretical and experiment accuracy one can test quantum structure of spacetime in particle colliders far below Planck scale. 
\section{Conclusions}
In this article we have recalculated the Higgs mass in the Standard Model by taking into account the gravitational corrections and asymptotic safety requirements using the current observational bounds on $\Mtop$. Due to the stability bound the Higgs mass is predicted to be a higher than the experimental value.\\
We have investigated the two beyond SM models which improve the running of $\lambda$. In the \textbf{Model I} we observed that with $\lambda_3 =0$ one gets $m_H \approx 125$ GeV as the lowest value, which agrees with the stability bound. However the other $Z'$ models can give different predictions. The correct Higgs mass can also be obtained in the Conformal Standard Model \cite{PhysRevD.99.115029} where a new scalar degree of freedom is also constrained to have $m_{\phi} \approx 300$ GeV. On the other hand we have excluded the possibility that the addition of sterile quarks gives the correct $m_H$.\\
Our analysis shows that the addition of new degrees of freedom can stabilise the electroweak vacuum at the experimental value of the Higgs Boson mass. One should stress once again that whole reasoning relies on the precise measurement of the top quark mass \cite{Bezrukov:2014ina} and hence the conclusions can be altered by future measurements.
\section*{Acknowledgments}
J.H.K. thanks Piotr Chankowski, Frederic Grabowski, Krzysztof Meissner and Mikhail Shaposhnikov for valuable and inspiring discussions and for comments on the first version of the article. J.H.K. also thanks anonymous referee for careful reading of the manuscript and constructive critique which resulted in substantially improved article. J.H.K. would like to acknowledge the Max Planck Institute for Gravitational Physics (Albert Einstein Institute) hospitality and support during this work. J.H.K. was 
 partially supported by the Polish National Science Center (NCN) grant DEC-2017/25/B/ST2/00165.
\addcontentsline{toc}{section}{The Bibliography}
\bibliography{mybibfile.bib}{}
\bibliographystyle{apsrev4-1}
\end{document}